\documentclass[a4paper]{jpconf}
\usepackage{graphicx}
\begin{document}
\title{Cosmic rays at the highest energies}

\author{Angela V. Olinto}

\address{Department of Astronomy and Astrophysics, 
Kavli Institute for Cosmological Physics, \\
The University of Chicago, Chicago, IL 60637, USA}

\ead{olinto@kicp.uchicago.edu}

\begin{abstract}
After a century of observations, we still do not know the origin of cosmic rays. I will review the current state of cosmic ray observations at the highest energies, and their
implications for proposed acceleration models and secondary astroparticle
fluxes. Possible sources have narrowed down with the confirmation of a GZK-like
spectral feature. The anisotropy observed by the Pierre Auger Observatory
may signal the dawn of particle astronomy raising hopes for high energy neutrino observations.
However, composition related measurements point to a different
interpretation. A clear resolution of this mystery calls for much larger statistics than the
reach of current observatories.
\end{abstract}

\section{Introduction}

The dominant component of cosmic rays observed on Earth is believed to originate in Galactic cosmic accelerators. A transition from Galactic to extragalactic cosmic rays should occur somewhere between  the {\it knee} in the cosmic ray spectrum at a PeV ($\equiv 10^{15}$ eV) and a few EeV ($\equiv 10^{18}$ eV). Above a few EeV,  the so-called ultrahigh energy cosmic rays (UHECRs) are most likely extragalactic. These are observed to reach energies that exceed $10^{20}$ eV posing some interesting and challenging questions: Where do they come from? How can they be accelerated to such high energies? What kind of particles are they? What is the spatial distribution of their sources? What do they tell us about these extreme cosmic accelerators? How strong are the magnetic fields that they traverse on their way to Earth? How do they interact with the cosmic background radiation? What secondary particles are produced from these interactions? What can we learn about particle interactions at these otherwise inaccessible energies? 

Below, we will briefly summarize the recent progress on the study of UHECRs. Recent reviews can be found in \cite{KKAO11,Letessier11}.
 
\section{Recent progress in UHECRs}

\subsection{Spectrum}

\begin{figure}[!t]
\centerline{\includegraphics[height=0.5\textheight]{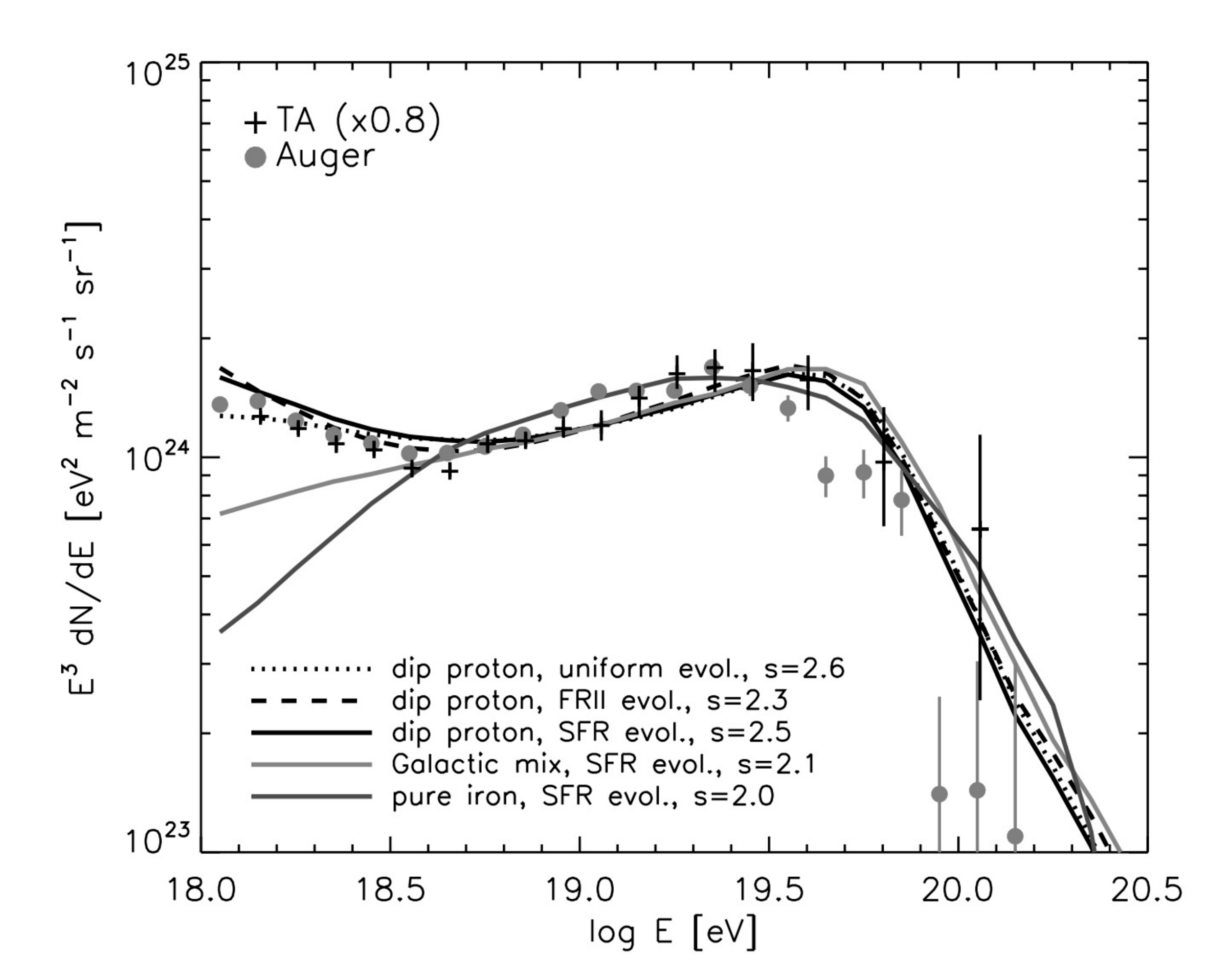}}
\caption{Flux of UHECRs multiplied by $E^3$ versus energy from the Auger Observatory \cite{AugerSpecICRC} and  the Telescope Array \cite{TATaup}. The TA absolute energy has been shifted
by multiplying by 0.8. The displayed error bars are statistical errors while the reported systematic error on the absolute energy scale is 22\%. Overlaid are simulated spectra obtained for different models of the Galactic to extragalactic transition and different injected chemical compositions and spectral indices, $s$ (adapted from \cite{KKAO11}). }
\label{uhecrSpec}
\end{figure}

Recent observations of UHECRs reveal a spectrum whose shape supports the long-held notion that sources of UHECRs are extragalactic. As shown in Figure \ref{uhecrSpec}, the crucial spectral feature recently established at the highest energies is a steeper decline in flux above about 30 EeV. This feature was first established by the HiRes Observatory \cite{Abbasi09} and confirmed with higher statistics by the Pierre Auger Observatory \cite{Abraham:2008ru}. This steep decline in flux is reminiscent of the effect of interactions between extragalactic cosmic rays and the cosmic background radiation, named the Greisen-Zatsepin-Kuzmin (GZK) cutoff \cite{G66,ZK66}, which causes cosmic ray protons above many tens of EeV to lose energy via pion photoproduction off cosmic backgrounds while cosmic ray nuclei photodissociate. This feature was not seen in earlier observations with the AGASA array \cite{Takeda98}. Data from the Auger Observatory \cite{AugerSpecICRC} and preliminary data from the Telescope Array \cite{TATaup} are shown in the  Figure  \ref{uhecrSpec}. The observations agree well given an overall energy re-scaling of 0.8 which is within the systematic errors in the absolute energy scale of 22\%.

Another important feature shown in Figure \ref{uhecrSpec}  is the hardening of the spectrum at a few EeV, called the {\it ankle}, which may be caused by the transition from Galactic to extragalactic cosmic rays or by propagation losses if UHECRs are mostly protons.

Figure \ref{uhecrSpec} shows the observed spectrum fit by different models of UHECR sources (adapted from \cite{KKAO11}). In the mixed composition and iron dominated models   \cite{Allard07}, the ankle indicates a transition from Galactic to extragalactic cosmic rays, the source evolution is similar to the star formation rate (SFR),  and the injection spectra are relatively hard (power law index $s\sim 2 -2.1$). In the proton dominated models in the figure, the ankle is due to pair production propagation losses \cite{BG88}, named ``dip transition models" \cite{BGG06}, and the injection spectra are softer for a wide range of evolution models. Models with proton primaries can also fit the spectrum with harder injection with a transition from Galactic to extragalactic at the ankle.

The confirmed presence of a spectral feature similar to the predicted GZK cutoff, settles the question of whether acceleration in extragalactic  sources can explain the high-energy spectrum, ending the need for exotic alternatives designed to avoid the GZK feature. However, the possibility that the observed softening of the spectrum is mainly due to the maximum energy of acceleration at the source, $E_{\rm max}$,  is not as easily dismissed. A confirmation that the observed softening {\it is} the GZK feature,  awaits supporting evidence from the spectral shape (at energies above 100 EeV), anisotropies (which are expected above GZK energies), composition, and the observation of produced secondaries such as neutrinos and photons.

\subsection{Anisotropies}

The landmark measurement of a flux suppression at the highest energies encourages the search for sources in the nearby extragalactic universe using the arrival directions of trans-GZK cosmic rays (with energy above $\sim$ 60 EeV). Above GZK energies, observable sources must lie within about 100 Mpc, the so-called GZK horizon or GZK sphere. At  trans-GZK energies, light composite nuclei are promptly dissociated by cosmic background photons, while protons and iron nuclei  may reach us from sources at distances up to about 100 Mpc. Since matter is known to be distributed inhomogeneously within this distance scale, the cosmic ray arrival directions should exhibit an anisotropic distribution above the GZK energy threshold, provided intervening magnetic fields are not too strong. At the highest energies, the isotropic diffuse flux from sources far beyond this GZK horizon should be strongly suppressed.

The most recent discussion of anisotropies in the sky distribution of UHECRs began with the report that the arrival directions of the 27 cosmic rays observed by Auger with energies above 57 EeV exhibited a statistically significant correlation with the anisotropically distributed galaxies in the 12th VCV \cite{VC06} catalog of active galactic nuclei (AGN)  \cite{Auger1,Auger2}. The correlation was most significant for AGN with redshifts $z <$ 0.018 (distances $< $ 75 Mpc)  and within 3.1$^\circ$ separation angles. An independent dataset confirmed the anisotropy at a confidence level of over 99\% \cite{Auger1,Auger2}. The prescription established by the Auger collaboration tested the departure from isotropy given the VCV AGN coverage of the sky, not the hypothesis that the VCV AGN were the actual UHECR sources. A recent update of the anisotropy tests with 69 events above 55 EeV \cite{Abreu10} shows that the correlation with the VCV catalog is not as strong for the same parameters as the original period (20 events correlate out of the original 27 while only 12 correlate out of the new 42). The data after the prescription period shows a departure from isotropy at the 3$\sigma$ level. In this meeting it was shown that of the 84 Auger events above 55 EeV observed after the original 14 events used to set up the prescription, 28 correlate \cite{AugerAnisTaup} which amounts to a (33 $\pm$ 5)\% correlation versus  21\% expected from isotropy. (Auger observed a total of 98 events above 55 EeV up to June 2011.) The Telescope Array showed that 8 out of 20 events correlate \cite{TATaup}, which is a 40\% correlation while  24\% is expected from isotropy. The two observations are consistent and show that an anisotropy signal is weak at these energies probably due to a large isotropic background. The lack of statistics at higher energies limits the reach of current observatories to achieve a clear detection if the anisotropy is due to the large scale structure and primaries are heavier than proton.

The anisotropy reported by the test with the VCV catalog  may indicate the effect of the large scale structure in the distribution of source harboring galaxies or it may be due to a nearby source. An interesting possibility is the cluster of Auger events around the direction of Centaurus A, the closest AGN (at $\sim$  3.8 Mpc). The most significant excess is in a 24 degree window around Cen A, where 19 of the 98 events are found, while  7.6 are expected by chance correlation  \cite{AugerAnisTaup}. The significance for the excess region can only be established with independent data. Only much higher statistics will tell if Cen A is the first UHECR source to be identified. 

\subsection{Composition}

The third key measurement that can help resolve the mystery behind the origin of UHECRs is their composition as a function of energy observed on Earth. Composition measurements can be made directly up to energies of $\sim$ 100 TeV with space-based experiments. For higher energies, composition is derived from the observed development and particle content of the extensive air shower created  by the primary cosmic ray when it interacts with the atmosphere. 

\begin{figure}[!t]
\centerline{\includegraphics[height=0.5\textheight]{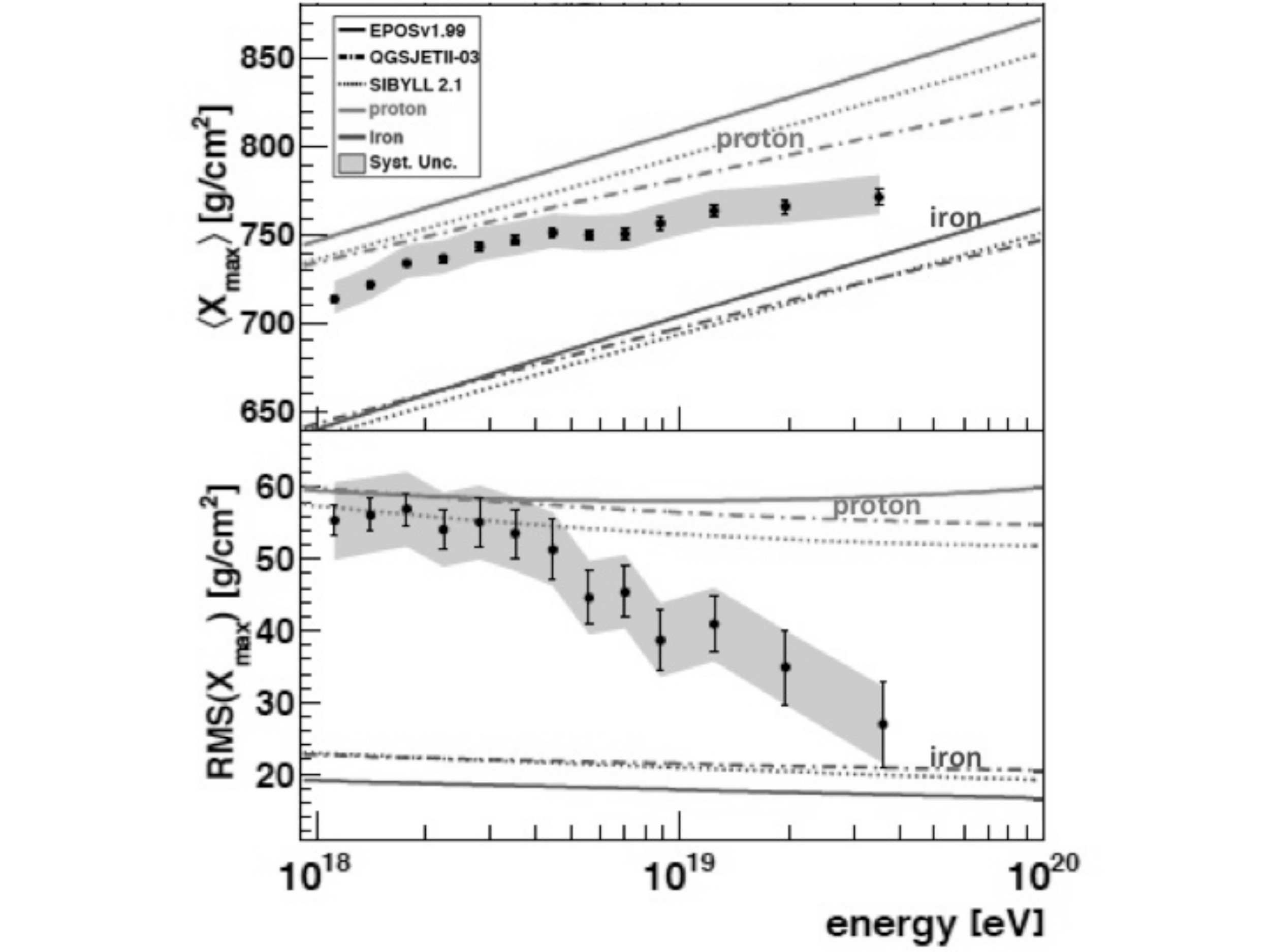}}
\caption{Average $X_{\rm max}$ (top panel) and the RMS of $X_{\rm max}$ (bottom panel)
are shown as a function of the energy. Auger data are the black points with statistical error bars. Systematic uncertainties are indicated as a grey band.
Predictions from different hadronic interaction models (EPOSv1.99 in solid lines, QGSJetII-03 in dash-dot lines,  and SIBYLL 2.1 in dotted lines)  for proton and iron primaries are shown as labelled. }
\label{AugerComp}
\end{figure}

Assuming that hadronic interactions models describe reasonably well the air shower properties of different primaries at these energies, observations show the dominance of light nuclei around a few EeV. As shown in Figure \ref{AugerComp}, a surprising trend occurs in data by the Auger Observatory above 10 EeV, a change toward heavy primaries is seen both in average position of the maximum of the showers  as well as in the RMS fluctuations about the mean up to 40 EeV \cite{AugerCompTaup,Abraham:2010yv}. As a mixture of different nuclei would increase the RMS fluctuations, the observed narrow distribution argues for a change toward a composition dominated by heavier nuclei. Using complementary techniques (asymmetry of the signal rise time and muon production depth) designed to make use of the high statistics of surface detector events, the Auger collaboration extended the measurement of shower properties to energies up to 60 EeV \cite{AugerCompTaup}. The trend toward heavier nuclei continues. The preliminary TA  measurement of fluctuations  remains closer to light primaries up to around 50 EeV. The two results are consistent within quoted errors, so the situation is currently unclear. 

As reported in this meeting, the study of showers of energies up to $10^{18.5}$ eV gives an estimate of the proton-air cross section of 505 mb $\pm 22$ (statistical uncertainty) and $+$ 28 $-$ 36 (systematic uncertainty) \cite{AugerHadrTaup}.  Changes to hadronic interactions from current extrapolations provide a plausible alternative interpretation to the observed shower development behavior above $10^{18.5}$ eV. Auger probes interactions above 100 TeV center of mass, while hadronic interactions are only known around a TeV. The observation of anisotropies and secondary particles (neutrinos and gamma-rays) can lead to astrophysical constraints on the composition of UHECRs, opening the possibility for the study of hadronic interaction cross sections, multiplicities, and other interaction parameters at hundreds of TeV.

The detailed composition of UHECRs is still to be understood, but it is clear that primaries are not dominated by photons \cite{Aglietta:2007,Abraham:2009qb} or neutrinos \cite{Auger_nu09,Abbasi08neu}. Limits on the photon fraction place stringent limits on models where UHECRs are generated by the decay of super heavy dark matter and topological defects. Unfortunately, the uncertainties on the UHECR source composition, spectrum, and redshift evolution translates to many orders of magnitude uncertainty in the expected cosmogenic neutrino flux as discussed next.

\subsection{Secondaries from UHECRs}

\begin{figure}[!t]
\centerline{\includegraphics[height=0.5\textheight]{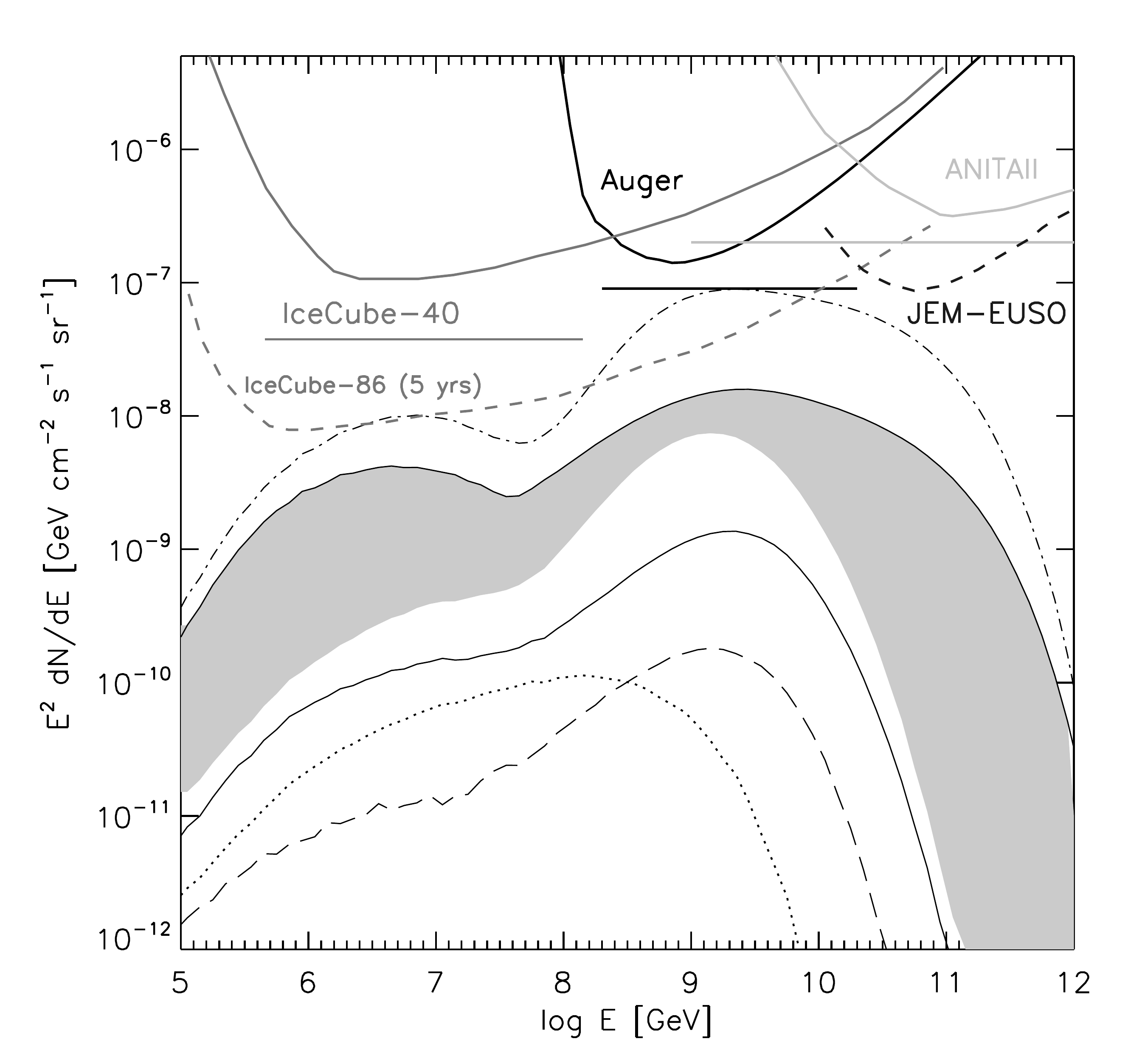}}
\caption{Cosmogenic neutrino flux for all flavors, for different UHECR parameters compared to instrument sensitivities (adapted from \cite{KAO10}). Dash-dotted line corresponds to a strong source evolution case (FRII evolution, see \cite{Wall05}) with a pure proton composition, dip transition model, and $E_{\rm max}=$ 3~ZeV. Uniform source evolution with: iron rich (30\%) composition and $E_{Z,\rm max}<Z$ 10 EeV is shown in the dotted line and the dashed line is for pure iron injection and $E_{Z,\rm max}=Z$ 100 EeV. Grey shaded range brackets dip and ankle transition models, with evolution of star formation history for $z<4$, pure proton and mixed `Galactic' compositions, and large proton $E_{\rm max} ( >  100$ EeV). Including the uniform source evolution would broaden the shaded area down to the black solid line. Experimental limits (solid lines) assume 90\% confidence level and full mixing neutrino oscillation. The differential limit and the integral flux limit on a pure $E^{-2}$ spectrum (straight line) are presented for IceCube 22 lines \cite{Abbasi10} and IceCube 40  \cite{Abbasi11}, ANITA-II \cite{ANITA10} and Auger \cite{Auger_nu09}. Dashed lines show future sensitivities for IceCube 80 lines \cite{Karle10}, and for JEM-EUSO  \cite{JemEUSO}.}
\label{cosmoNeut}
\end{figure}

Secondary neutrinos and photons can be produced by UHECRs when they interact with ambient baryonic matter and radiation fields inside the source or during their propagation from source to Earth. These particles travel in geodesics unaffected by magnetic fields and bear valuable information of the birthplace of their progenitors. The quest for sources of UHECRs has thus long been associated with the detection of neutrinos and gamma rays that might pinpoint the position of the accelerators in the sky.

The detection of these secondary particles is not straightforward however: first, the propagation of gamma rays with energy exceeding several TeV is affected by their interaction with CMB and radio photons. These interactions lead to the production of high energy electron and positron pairs which in turn up-scatter CMB or radio photons by inverse Compton processes, initiating electromagnetic cascades. As a consequence, one does not expect to observe gamma rays of energy above $\sim 100$~TeV from sources located beyond a horizon of a few Mpc. Above EeV energies, photons can again propagate over large distances, depending on the radio background, and can reach observable levels around tens of EeV. Secondary neutrinos are very useful because, unlike cosmic-rays and photons, they are not absorbed by the cosmic backgrounds while propagating through the Universe. In particular, they give a unique access to observing sources at PeV energies. However, their small interaction cross-section makes it difficult to detect them on the Earth requiring the construction of km$^3$ or larger detectors.

Neutrinos generated during UHECR propagation \cite{BZ69,Stecker79}, often called cosmogenic neutrinos, represent a ``guaranteed flux'' and have encouraged efforts to detect them for decades (see, e.g.,  \cite{AM09}). One important assumption for the existence of cosmogenic neutrinos, that cosmic rays are extragalactic at the highest energies, has been verified by the detection of a feature consistent with the GZK cutoff  in the cosmic ray spectrum \cite{Abbasi09,Abraham:2008ru}  and by the indication of anisotropies in the cosmic ray sky distribution at the highest energies  \cite{Auger1,Auger2}. 
These findings herald a possible resolution to the mystery behind the origin of UHECRs and the possibility of detecting ultrahigh energy neutrinos in the near future. 

This optimistic view has been dampened by the indication that UHECRs may be dominated by heavier nuclei \cite{AugerCompTaup,Abraham:2010yv}. The cosmogenic neutrino flux expected from heavy cosmic ray primaries can be much lower than if the primaries are protons at ultrahigh energies, making a detection extremely challenging for current observatories.  Conversely, if neutrinos are observed, they will test specific sets of cosmic ray source parameters.

Figure~\ref{cosmoNeut} summarizes the effects of different assumptions about the UHECR source evolution, the Galactic to extragalactic transition, the injected chemical composition, and $E_{\rm max}$, on the cosmogenic neutrino flux (adapted from \cite{KAO10}). It demonstrates that the parameter space is  poorly constrained with uncertainties of several orders of magnitude in the predicted flux.

Due to the delay induced by cosmic magnetic fields on charged cosmic rays, secondary neutrinos and photons should not be detected in time coincidence with UHECRs if the sources are not continuously emitting particles, but are transient such as gamma-ray bursts and young pulsars.

\section{Conclusion}

The resolution of the long standing mystery of the origin of ultrahigh energy cosmic rays will require a coordinated approach on three complementary fronts: the direct ultrahigh energy cosmic ray frontier, the transition region between the knee and the ankle, and the multi-messenger interface with high-energy photons and neutrinos.

\begin{figure}[!t]
\centerline{\includegraphics[height=0.5\textheight]{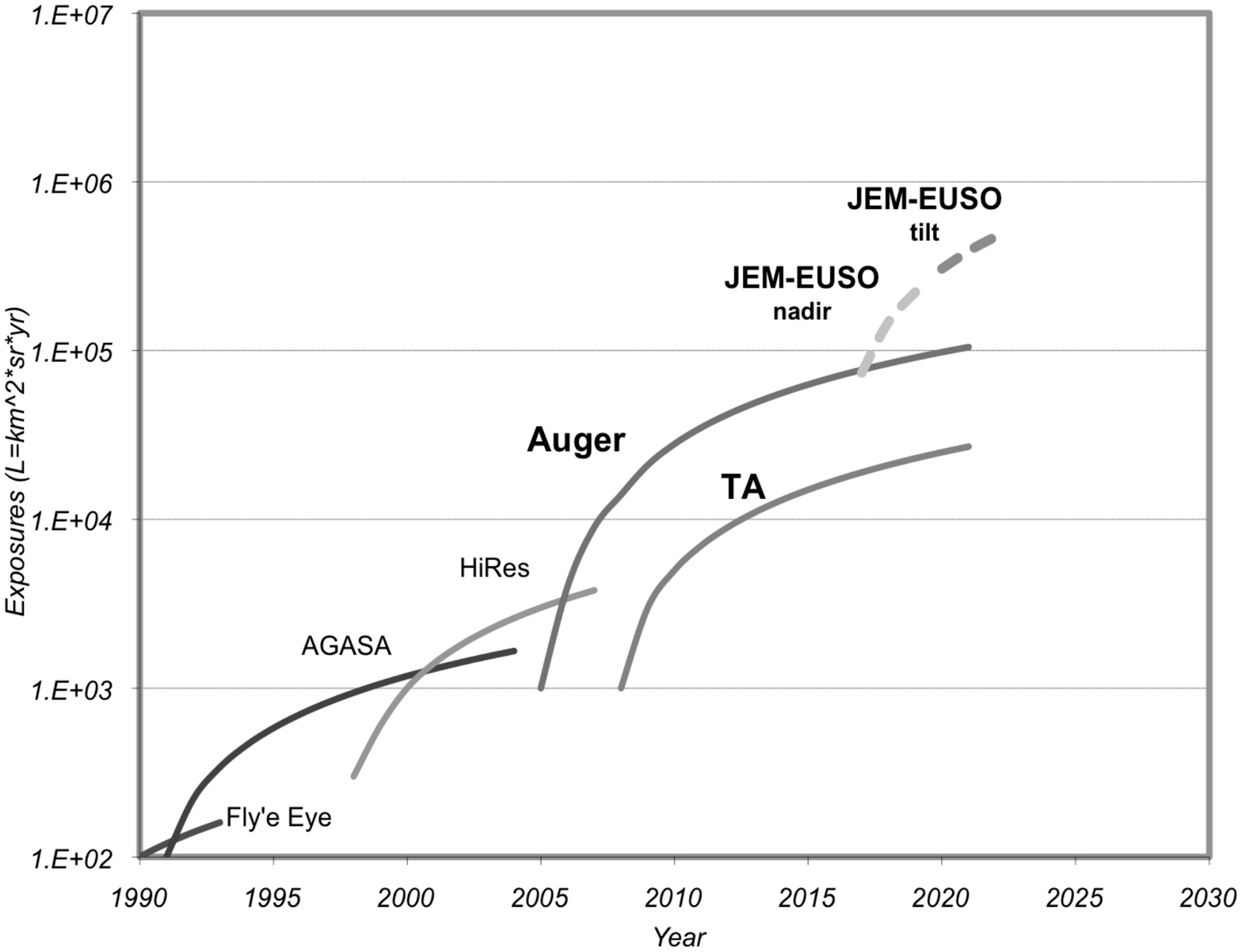}}
\caption{Exposures to UHECRs from 1990 to present from Fly's Eye, AGASA, HiRes, Auger and TA. Estimates of the exposures in the future of Auger and TA plus the planned space observatory,  JEM-EUSO \cite{JemEUSO}.}
\label{ExposFutur}
\end{figure}

Current data suggest that watershed anisotropies will only become clear above 60 EeV and that very large statistics with good angular and energy resolution will be required. The Auger Observatory (located in Mendoza, Argentina),  will add $7\times10^3 $ km$^2$ sr each year of exposure to the southern sky, while the Telescope Array (located in Utah, USA) will add about $2\times10^3 $ km$^2$ each year in the North as shown in Figure \ref{ExposFutur}. Current technologies can reach a goal of another order of magnitude if deployed by bold scientists over very large areas. New technologies may ease the need for large number of detector units to cover similarly large areas. 

A promising avenue to reach the necessary high statistics is the idea of space observatories (e.g., JEM-EUSO, OWL, Super-EUSO). With current technologies, a large statistics measurement of the spectrum and angular distribution of arrival directions above GZK energies are well within reach. 
Improved photon detection technologies will be needed to reconstruct shower maxima from space. If deployed in 2017, JEM-EUSO can significantly increase the exposure to UHECRs reaching the level needed to unveil this mystery \cite{JemEUSO} as  in Figure \ref{ExposFutur}.

With a coordinated effort, the next generation observatories can explore more of the $\sim 5$~million trans-GZK events the Earth's atmosphere receives per year and find the highest energy accelerators in the universe.

\ack
Thanks to Kumiko Kotera and the Auger group  for very fruitful discussions. This work was supported by the NSF grant PHY-1068696 at the University of Chicago, and the Kavli Institute for Cosmological Physics through grant NSF PHY-1125897 and an endowment from the Kavli Foundation.


\section*{References}

\begin{thebibliography}{99}

\bibitem{KKAO11}
K.~Kotera, A.~V. Olinto, Annu. Rev. Astron. Astrophys. 49 (2011) 119-153.

\bibitem{Letessier11}
A.~{Letessier-Selvon}, T.~{Stanev}, Rev. Mod. Phys. 83 (2011) 907.

\bibitem{Abbasi09}
R.~U. {Abbasi}, et~al., Astropart. Phys. 32 (2009) 53.

\bibitem{Abraham:2008ru}
J.~{Abraham}, et~al., Phys. Rev. Lett. 101 (2008) 061101.

\bibitem{G66}
K.~Greisen, Phys. Rev. Lett. 16 (1966) 748.

\bibitem{ZK66}
G.~Zatsepin, V.~Kuzmin,  J. Exp. Theor. Phys. Lett. 4 (1966) 78.


\bibitem{AugerSpecICRC}
F. Salamida for the Pierre Auger Collaboration, Proceed. of the 32nd ICRC (2011), Beijing, China.

\bibitem{TATaup}
D. Ikeda for the Telescope Array Collaboration, in these Proceedings

\bibitem{Takeda98}
M.~{Takeda}, et~al., Phys. Rev. Lett. 81 (1998)  1163.

\bibitem{Allard07}
D.~{Allard}, E.~{Parizot}, A.~V. {Olinto}, Astropart. Phys. 27 (2007) 61.

\bibitem{BG88}
V.~S. {Berezinsky}, S.~I. {Grigorieva},  A\&A 199 (1988) 1.

\bibitem{BGG06}
V.~{Berezinsky}, A.~{Gazizov}, S.~{Grigorieva}, Phys. Rev. D 74 (2006) 043005.

\bibitem{VC06}
M.-P. {V{\'e}ron-Cetty}, P.~{V{\'e}ron},  A\&A 455 (2006) 773.

\bibitem{Auger1}
J.~{Abraham}, et~al.,  Science 318 (2007) 938.

\bibitem{Auger2}
J.~{Abraham}, et~al.,  Astropart. Phys. 29 (2008)  188.

\bibitem{Abreu10}
P.~{Abreu}, et~al.,  Astropart. Phys. 34 (2010) 314.

\bibitem{AugerAnisTaup} 
C. Macolino for the Pierre Auger Collaboration, in these Proceedings


\bibitem{AugerCompTaup} 
L. Cazon for the Pierre Auger Collaboration, in these Proceedings

\bibitem{Abraham:2010yv}
J.~{Abraham}, et~al., Phys. Rev. Lett. 104 (2010) 091101.

\bibitem{AugerHadrTaup}
R. Ulrich for the Pierre Auger Collaboration, in these Proceedings

\bibitem{Aglietta:2007}
J.~Abraham, et~al.,  Astropart.  Phys. 29 (2008) 243.

\bibitem{Abraham:2009qb}
J.~{Abraham}, et~al., Astropart. Phys. 31 (2009)  399.

\bibitem{Auger_nu09}
J.~{Abraham}, et~al., Phys. Rev. D 79 (2009) 102001.

\bibitem{Abbasi08neu}
R.~U. {Abbasi}, et~al., Astrophys. J. 684 (2008) 790.

\bibitem{BZ69}
V.~S. {Berezinsky}, G.~T. {Zatsepin}, Phys. Lett. B 28 (1969) 423.

\bibitem{Stecker79}
F.~W. {Stecker},  Astrophys. J.  228 (1979) 919.

\bibitem{AM09}
L.~A. {Anchordoqui}, T.~{Montaruli}, Ann. Rev. Nucl. Part. Sci. 60 (2010) 129.

\bibitem{KAO10}
K.~{Kotera}, D.~{Allard}, A.~V. {Olinto}, JCAP 10 (2010) 13.

\bibitem{Wall05}
J.~V. {Wall}, C.~A. {Jackson}, P.~A. {Shaver}, I.~M. {Hook}, K.~I.
 {Kellermann},  A\&A 434 (2005) 133.

\bibitem{Abbasi10}
R.~U. {Abbasi}, et~al., Phys. Rev. Lett. 104 (2010) 161101.

\bibitem{Abbasi11}
R.~U. {Abbasi}, et~al., Phys. Rev. D 84 (2011) 082001.

\bibitem{ANITA10}
P.~W. {Gorham}, et~al.,  Phys. Rev. D 82 (2010) 022004.

\bibitem{Karle10}
A.~{Karle}, {IceCube}, Proceedings of the 31st ICRC, Lodz, Poland (2009), arXiv:1003.5715.

\bibitem{JemEUSO}
Y.~{Takahashi}, et~al., {The Jem-Euso Mission}, New J. Phys. 11 (2009) 065009.



\end{thebibliography}
\end{document}